\documentclass[pra,
                twocolumn,
                preprintnumbers,
                amsmath,
                amssymb,
                showpacs,
                superscriptaddress]{revtex4-1}
\usepackage{graphicx} % Include figure files
\usepackage{dcolumn}  % Align table columns on decimal point
\usepackage{bm}       % bold math
\usepackage{amsfonts}
\usepackage{amssymb}
\usepackage[dvipsnames]{xcolor}
\usepackage{soul}
\usepackage{braket}
\usepackage{amsmath}
\usepackage[normalem]{ulem}

\newcommand{\eq}[1]{Eq.~(\ref{#1})}
\newcommand{\ful}{\mbox{C$_{60}$}}

\begin{document}

\title{Ultrafast nonadiabatic electron dynamics in photoexcited $\ful$: A comparative study among DFT exchange-correlation functionals}

\author{Esam Ali}
\affiliation{Department of Natural Sciences, Dean L. Hubbard Center for Innovation, Northwest Missouri State University, Maryville, Missouri
64468, USA}
\affiliation{Department of Physics, Faculty of Science, University of Benghazi, Benghazi 9480, Libya}

\author{Mohamed El-Amine Madjet}
\affiliation{Department of Natural Sciences, Dean L. Hubbard Center for Innovation, Northwest Missouri State University, Maryville, Missouri 64468, USA}
\affiliation{Bremen Center for Computational Materials Science, University of Bremen, 28359 Bremen, Germany}

\author{Ruma De}
\affiliation{Department of Natural Sciences, Dean L. Hubbard Center for Innovation, Northwest Missouri State University, Maryville, Missouri
64468, USA}

\author{Thomas Frauenheim}
\affiliation{Constructor University, Campus Ring 1, 28759 Bremen, Germany}
\affiliation{Beijing Computational Science Research Center, 100193 Beijing, China}
\affiliation{Shenzhen JL Computational Science and Applied Research Institute, 518110 Shenzhen, China}

\author{Himadri S. Chakraborty}
\email{himadri@nwmissouri.edu}
\affiliation{Department of Natural Sciences, Dean L. Hubbard Center for Innovation, Northwest Missouri State University, Maryville, Missouri
64468, USA}

\begin{abstract}
{The non-radiative electron-relaxation dynamics in $\ful$ molecule is studied after selective initial photoexcitations. The methodology includes nonadibabtic molecular simulation combined with time-dependent density functional theory (DFT) and semi-classical surface hopping approach. Results of treating the DFT exchange-correlation (xc) interaction by the non-empirical Perdew-Burke-Ernzerhof (PBE), hybrid PBE0, and hybrid Becke 3-parameter Lee–Yang–Parr (B3LYP) functional are compared. Even though some differences in the details are found, all three functionals produce qualitatively similar unoccupied band structures in the ground state. The model-dependent differences in the ultrafast population dynamics, including the occurrences of transient entrapment of population, are studied systematically. The trend of the results demonstrates a universal dependence on the structure of unoccupied band offering a spectroscopic route to probe this structure. Results can be verified, as well as the best xc model for quantitative accuracy can be determined, by comparing with ultrafast transient absorption or time-resolved photoelectron spectroscopy measurements. From the computational standpoint, the study facilitates method optimization to simulate nonadiabatic relaxation dynamics in technologically important fullerene derivatives.}

\end{abstract}

\maketitle

\section{Introduction}

%%% Broader applications of fullerene-based materials
Trends of technology utilize fullerene-based materials, due to their stability and unique properties, through applications in energy storage~\cite{friedl18} and conversion\cite{jeon19,collavini18}. For instance, pump-probe ultrafast transient absorption spectroscopy (UTAS) measurements of blends of a low-band-gap polymer with fullerene derivatives found more efficient charge generation process versus in pristine polymer~\cite{vezie2019}. UTAS has also been applied to access effects of subtle cage-symmetry in visible-light driven electron dynamics of metallofullerenes which is of interest in visible-light solar energy harvesting~\cite{wu2015}. Likewise, to aid ultraviolet (UV) harvesting for solar cell design, the UTAS approach was undertaken to study charge transport and electron transfer in a PbI$_2$/$\ful$ heterojunction that showed a higher charge extraction efficiency~\cite{cheng2019}. On the other hand, two-photon time-resolved photoelectron spectroscopy (TRPES) is applied to disclose optical generation of non-interacting excitons in a fullerene film leading to redistribution of transport levels of the non-excited molecules~\cite{emmerich21}. Therefore, ultrafast dynamics of charge separation, migration, transport, transient trap, and recombination of photoexcited electrons in these materials are important processes. 

%%% Charge transfer process in general photovoltaics
Charge transfer (CT) is the key sub-process that underpins the core mechanism of organic photovoltaics.  The donor-acceptor complexes in this application are abundantly composed of fullerene materials. A fullerene molecule can be structurally upgraded to attend desired chemical properties in general. This can be achieved by, namely, choosing its endohedral species~\cite{ross09} and/or exohedral ligands or polymers~\cite{he11,li12}. The objective is to control the light absorption efficiency and carrier transport. An excitonic state is created upon absorbing a photon by the complex. This exciton either dissociates into free carriers or recombines to decay depending on the dominance of, respectively, the electron-hole separation energy or excitonic binding energy. The dissociation is the preferred mechanism for photovoltaics~\cite{emmerich21}, the probability of which may enhance as the decay times of excited electrons elongate. Therefore, the decay and transfer of a hot electron from one location of the molecular material to another is a fundamental sub-process of these events~\cite{sato18,ortiz17,boschetto18,juvenal19,cheng19}. Conversely, a relatively quicker relaxation may favor re-population of cold electrons and subsequent thermalization of the molecular lattice. The latter process has applications in photothermal cancer therapy. In fact, a high photothermal efficiency and superior stability is found in polyhydroxy fullerenes for it to become an ideal candidate for such applications~\cite{chen2020}. 

%%% Nonadiabatic relaxation in fullerenes
Since fullerene molecules constitute the key moiety in these compounds, the understanding of their CT processes will greatly benefit by investigating electron relaxation in a pristine fullerene molecule. Besides, there is other fundamental importance. Experimentally, UTAS~\cite{berera09,bhattacherjee2017,gesuele19} and TRPES~\cite{emmerich21,stadtmueller19}, using fs pulses or, more recently, attosecond pulses for greater resolution~\cite{driver20,geneaux19}, can probe such dynamics in real time. Indeed, photoinduced charge migration has been measured in the time domain for fullerene-based polymerized films~\cite{juvenal19} and heterojunctions~\cite{cheng19}, and also for bulks~\cite{he18} and nanorods~\cite{kedawat19}. Pristine $\ful$ is relatively easily available to conduct precision measurements. These ultrafast processes occur on the femtosecond (fs) time-scale and are driven by the strong coupling between ionic and electronic degrees of freedom. Therefore, computationally, frameworks based on nonadiabatic molecular dynamics (NAMD) are appropriate for providing accurate, comprehensive descriptions of the processes~\cite{sato18,nelson20}. The NAMD results presented in this study are obtained employing three variants of exchange-correlation (xc) functional within the scope of density functional theory (DFT). Comparisons with future measurements will examine the relative accuracy of the models. Resulting knowledge and understanding will facilitate the development and optimization of our computational methodology to address complex systems.  

\section{Description of Methods}

The software packages and the computation work-flow follow our previous study of the Mg@$\ful$~\cite{madjet2021} molecule, and references therein. Gamess-US~\cite{gamess1,gamess2} is used for the ground-state geometry optimization of $\ful$ conducted at the B3LYP/6-311+G$^{**}$ level of theory. This produced a good description of the band gap, 2.72 eV, close to reference values~\cite{vinit17,zhang14}. Moreover, the calculated difference of 5.15 eV between the $\ful$ ionization energy and electron affinity closely agreed with the difference of these quantities measured, respectively, by electron impact mass spectrometry~\cite{muigg96} and high-resolution photoelectron imaging~\cite{huang14}. 

%From the optimized structure of $\ful$, the molecular dynamics (MD) simulations for 6000 steps with a 0.5-fs step-size are conducted in the number-volume-temperature (NVT) canonical ensemble at 300 K employing a velocity-rescaling thermostat to maintain the temperature.  A production run is then conducted in the number-volume-energy (NVE) ensemble that extends to another 3000 fs in 0.5 fs steps. All these MD simulations are performed using the CP2K~\cite{cp2k} and deploying three separate choices of exchange-correlation functional: (i) non-empirical Perdew-Burke-Ernzerhof (PBE), (ii) hybrid PBE0,  and (iii) hybrid Becke 3-parameter Lee–Yang–Parr (B3LYP). These three tracks of xc functional were maintained up to the final results. 
The molecular dynamics (MD) simulations were conducted on the optimized structure of $\ful$ for 6000 steps with a step-size of 0.5 fs. The simulations were performed in the NVT canonical ensemble at 300 K using a velocity-rescaling thermostat to maintain the temperature. Subsequently, a production run was performed in the NVE canonical ensemble, extending for another 3000 fs with 0.5 fs a step. All MD simulations were carried out using the CP2K software~\cite{cp2k}, and three different choices of exchange-correlation functional were employed: (i) non-empirical (non-hybrid) Perdew-Burke-Ernzerhof (PBE), (ii) hybrid PBE0, and (iii) hybrid Becke 3-parameter Lee-Yang-Parr (B3LYP). These functionals were used throughout the simulations until the corresponding final results were obtained. The time-dependent populations were obtained by averaging over 20 initial configurations and 1000 stochastic realizations of surface hopping trajectories were performed for each configuration. The DFT-D3 dispersion correction of Grimme is used to account for the dispersion interactions~\cite{Grimme1,Grimme2}. The QMflows-namd module~\cite{qmflows} interfaced with CP2K is employed to compute electronic structure properties of molecular orbitals, energies and electron-phonon nonadiabatic couplings (NACs) between orbitals to construct the vibrionic Hamiltonian; see Subsection III\,C for more details. The early thermal equilibration phase of NVT is important to obtain accurate NACs. The energies and NACs are then used to perform the NAMD simulations using the PYXAID package; for details, see Ref.\,\cite{akimov1,akimov2,madjetccp,madjetjpcl}. Note that the isosurface plots in Figs.\,1 and 4 are obtained from the last step of the NVT simulation. 

Obviously, the methodology relies on an effective independent particle framework. Our previous study~\cite{madjet2021} employing a configuration-interaction description of many-body effects suggested that the many-body dynamics, which dominates the plasmon-driven ionization spectra~\cite{madjet08} at extreme UV (XUV) energies around 20 eV, is not critical in the near to far UV energy region of current interest. Dynamics in these energy regions are predominantly driven by electron-phonon coupling.
%%%
\begin{figure}[h!]
%\centering
\includegraphics[width=8.5 cm]{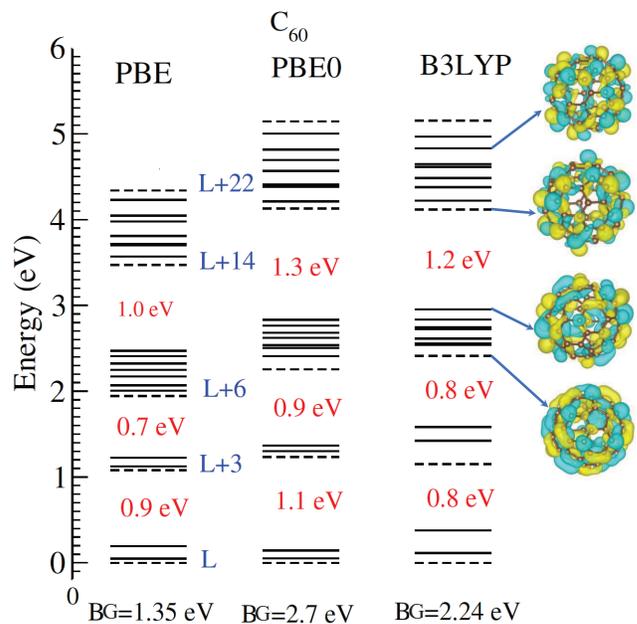}
\caption{(Color online) (a): $\ful$ unoccupied molecular orbital energies, relative to LUMO (L) up to LUMO+22 calculated at the DFT level of theory with the PBE, PBE0 and B3LYP exchange-correlation functional. The HOMO-LUMO band gap (BG) energy determined by each method is noted. Other energy gaps (values quoted) inlay the structures which are qualitatively identical for the methods while quantitatively somewhat different. DFT/B3LYP generated isosurface plots of LUMO+20, LUMO+14, LUMO+13, and LUMO+6 orbitals are illustrated. All information in this figure is taken from the last step of NVT simulation.}
\label{fig1}
\end{figure}
%%%

\section{Results and Discussions}

\subsection{LUMO Structures}

Fig.\,\ref{fig1} displays the unoccupied molecular orbital energies, referenced from the lowest unoccupied molecular orbital (LUMO) level, of $\ful$ up to LUMO+22 calculated using the PBE, PBE0 and B3LYP functional. PBE0 and B3LYP functionals are seen to yield almost identical energy structures. Non-hybrid PBE energies, on the other hand, are quantitatively somewhat compressed overall, but the structure is qualitatively quite similar. The HOMO-LUMO band gap is smaller on PBE compared to its values in PBE0 and B3LYP which are roughly identical. Also notice the three energy gaps that open up below LUMO+3, LUMO+6 and LUMO+14 in all three sets indicating a broad universality of the results irrespective of the choice of a xc functional. We note that such intermittent gaps in $\ful$ unoccupied levels were found in other calculations as well~\cite{madjet2021,schmidt15}. These gaps will produce strong transient events in the dynamics that we will reveal below. Electrons from, namely, the highest occupied molecular orbital (HOMO) level can be conveniently photoexcited by UV pump pulses across the HOMO-LUMO band gap to selected LUMO+$n$ excited states. These excited states are treated as the initial states in our simulation. Non-radiative population-decay, driven by electron-phonon couplings, of these states then becomes the dominant decay process. This process will eventually terminate by recombining with the HOMO vacancy. Note that no Auger type decay channel exists in this process, since the excitation energy is lower than the first ionization threshold of $\ful$.  
%%%
\begin{figure}[h!]
%\centering
\includegraphics[width=9.0 cm]{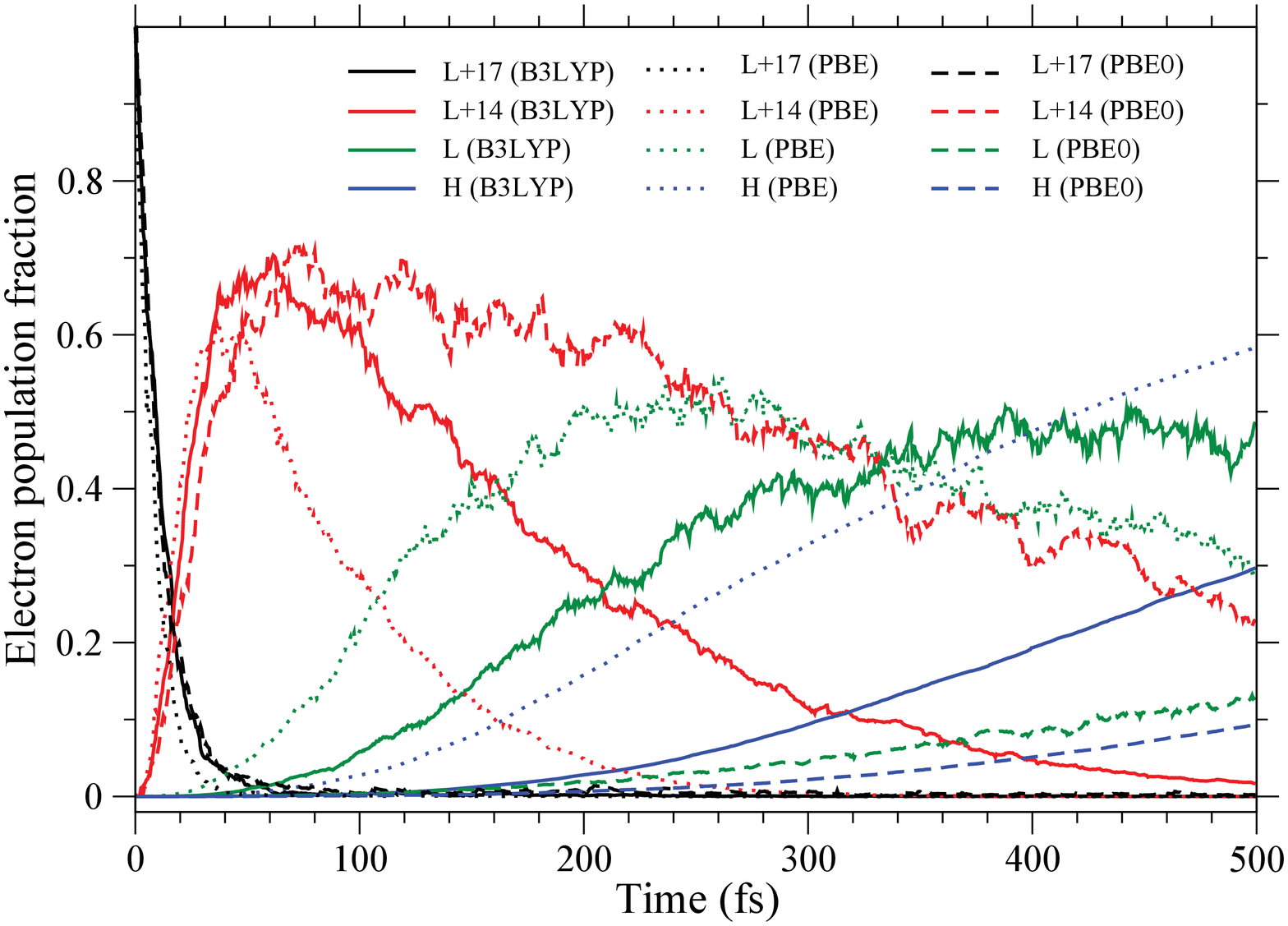}
%\vskip 1.5 cm
\includegraphics[width=9.0 cm]{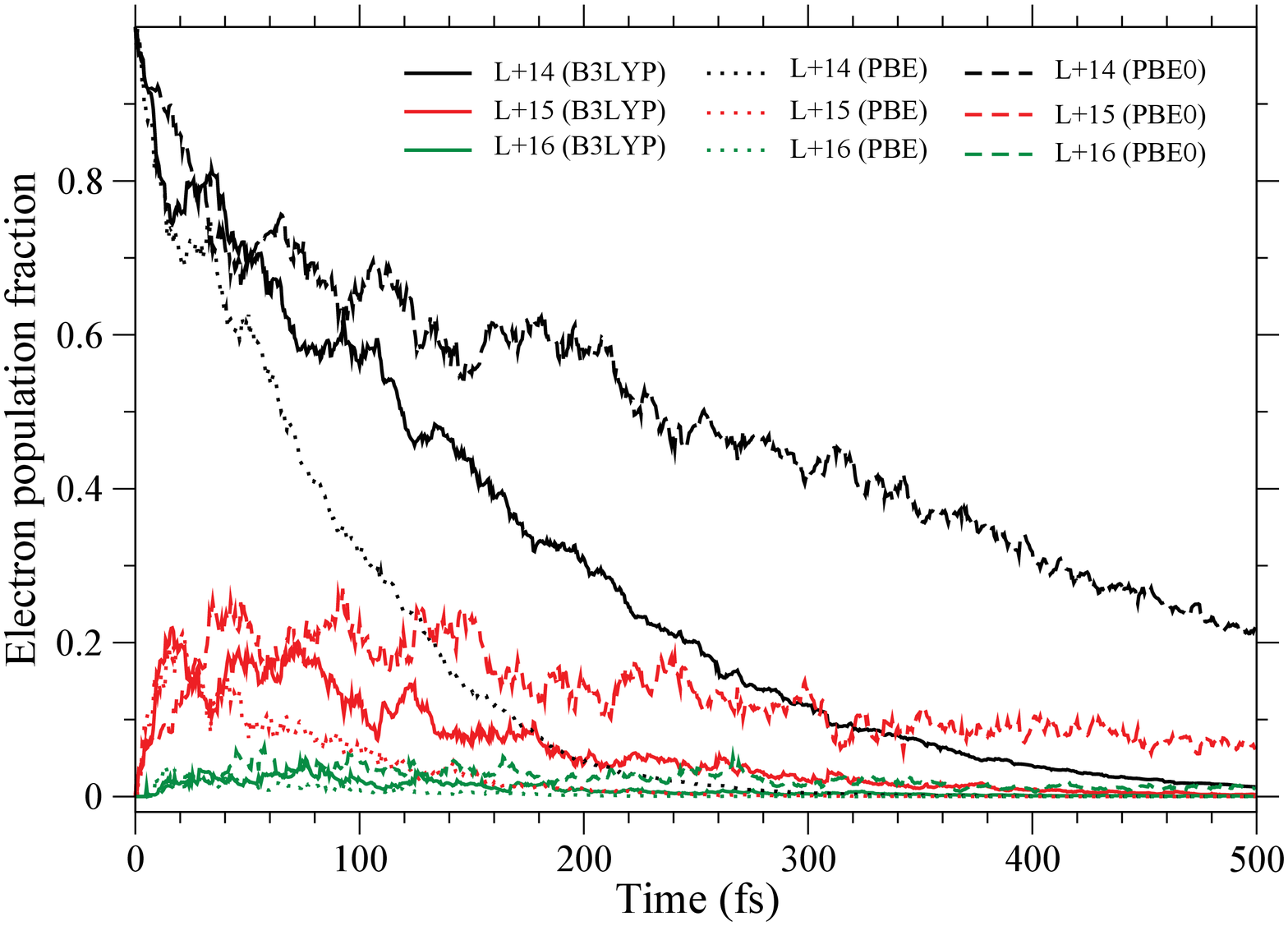}
\caption{Top: Time evolutions of the population of initially excited LUMO+17 and intermediate LUMO+14 state, and the partial evolution of LUMO and HOMO (recombination) state calculated using the B3LYP, PBE and PBE0 functional. Bottom: Same as top, but for the initial LUMO+14 state and subsequently sputtered population to higher LUMO+15 and LUMO+16 states.}
\label{fig2}
\end{figure}
%%%

\subsection{Ultrafast Evolution of Populations Dynamics}

The top panel of Fig.\,\ref{fig2} displays the simulated population time-evolution of the initially excited (100\% population at time zero) LUMO+17 state in three xc functionals. 
%The small oscillations in time correspond to the vibrational motion on the ground state potential energy surface superimposed with the electronic dynamics; this can be made somewhat smoothed with larger samplings of initial conditions. In any case, 
As seen, LUMO+17 depopulates in roughly 50 fs in all three treatments, while PBE exhibits the fastest decay. As the electron decays to LUMO+14, it experiences a transient entrapment due to a wide energy-gap below this state which slows down its subsequent decay. This effect is found universal in three xc approaches. The population dynamics of any intermediate state involves a combination of a growth and a decay dynamics. As a result, for LUMO+14, the growth due to the electron transfer from higher states dominates during earlier times to peak the cumulative population to 60\% or above. Subsequently, the decay begins to dominate and continues over remarkably longer times. Notably, PBE0 yields the highest peak population which also experiences the longest decay period. B3LYP closely follows up to the peaking time, although later decaying relatively faster. On the other hand, PBE features a slightly smaller maximum population with a significantly fast decay trail. These details are partly a direct ramification of differences in the size of the gap immediately below LUMO+14 in three xc approaches (Fig.\,\ref{fig1}). However, there are more in this which will be discussed in Subsection III\,D. The similar comparative trend in the dynamics of LUMO at the band-edge as well as the return of population to HOMO (recombination) among the results of three xc functionals is also seen in this figure. It may be noted that the peak population growth is not captured for LUMO in B3LYP and for HOMO in all three methods within the displayed range up to 500 fs.
%%%
\begin{figure*}[t]
%\centering
\includegraphics[width=0.9\textwidth]{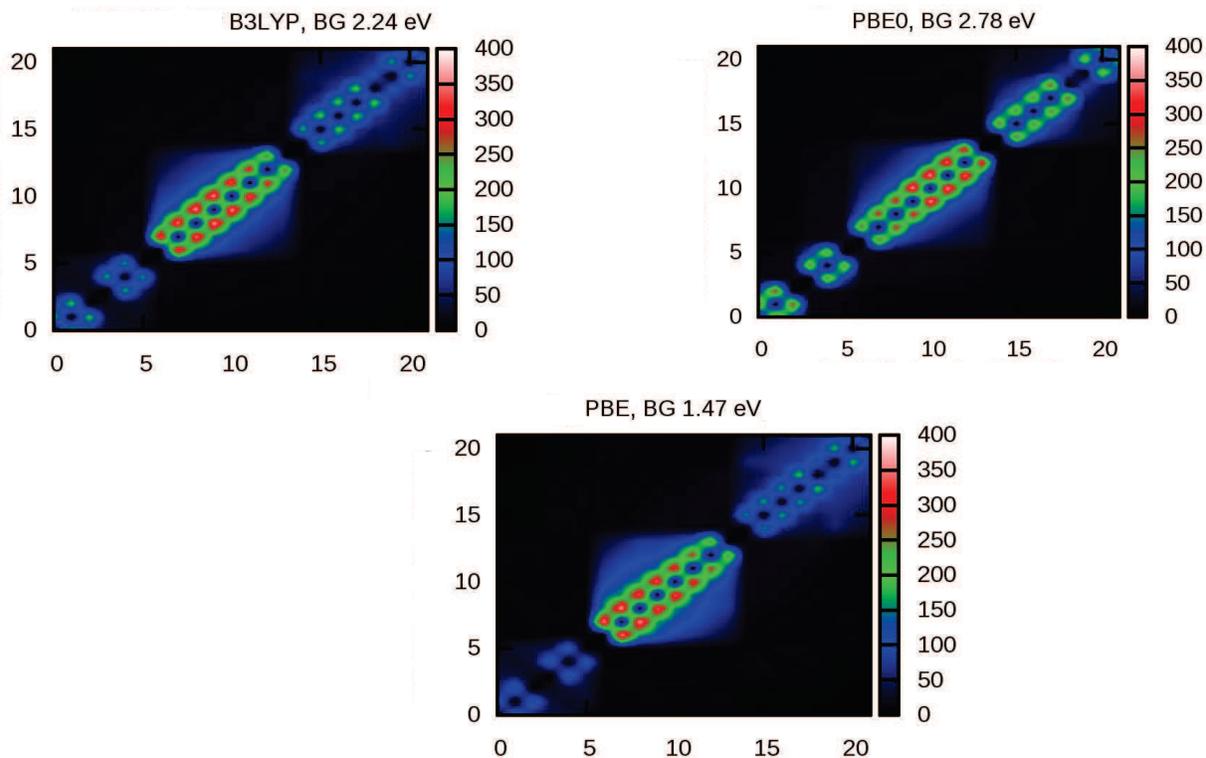}
\caption{(Color online) Magnitudes of MD-averaged nonadiabatic couplings (NACs) calculated using the B3LYP, PBE0 and PBE functional. The axes plot the index $n$ of LUMO+$n$ states. The band gap (BG) values are found after the NVE simulation and thus slightly differ from the values quoted in Fig.\,\ref{fig1}.}
\label{fig3}
\end{figure*}
%%%

To further examine the procrastinated decay of the LUMO+14 state, we present its dynamics in the bottom panel of Fig.\,\ref{fig2} by choosing this state to be the initially excited state. As seen, the general comparative temporal behavior among the methods discussed above is clearly retained. That is, PBE0 induces the slowest decay, while PBE is the fastest. This figure further discloses another feature of the mechanism that contributes to the dynamics. This entails some population of LUMO+14 to be {\em promoted} to LUMO+15 and even to LUMO+16. This highlights the role of the nuclear vibration-driven electron dynamics within the quasi-degenerate, compact states. As seen, due to the longer holdup of electrons in LUMO+14, some of the probability transitions higher in energy to LUMO+15 to populate it up to a maximum of about 20\% both in PBE0 and B3LYP, while to a lesser extent in PBE. A similar behavior is seen for LUMO+16 as well but in a much smaller scale. Obviously, this ``sputtered" electron-density eventually returns to {\em repopulate} LUMO+14 -- an effect that further favors LUMO+14's delayed decay. An identical behavior of the $\ful$ LUMO+14 level was earlier found in our study of Mg to $\ful$ ultrafast CT relaxation in the endohedral Mg@$\ful$~\cite{madjet2021}. Note further that the decay of the population promoted to LUMO+15 is slow as well due to the congestion caused by the energy gap, although it readily repopulates LUMO+14.

\subsection{Nonadiabatic Couplings}

The electronic wavefunction of the molecule can be expressed by a time-dependent linear superposition of molecular orbitals $\phi_j$ with, say, $C_j$ being the mixing coefficients that act as the electronic degrees of freedom. $C_j$ must evolve by the time-dependent Scr\"{o}dinger equation in natural units as,
\begin{equation}\label{tdse}
i\frac{\partial}{\partial{t}} C_j(t) = \sum_{k} H^{\scriptsize{\mbox{vib}}}_{jk}(\vec{R}(t))C_k(t).
\end{equation}
Here the vibrionic Hamiltonian matrix $H^{\scriptsize{\mbox{vib}}}$ can be written as,
\begin{equation}\label{vib-ham}
H^{\scriptsize{\mbox{vib}}}_{jk} = \epsilon_j(\vec{R}(t))\delta_{jk} - id_{jk}(\vec{R}(t)),
\end{equation}
where $\vec{R}$ is the nuclear coordinate, and $\delta_{jk}$ are Kronecker delta symbols.
The computed $\phi_j$ and orbital energies $\epsilon_j$ along the nuclei trajectories are used to obtain NACs $d_{jk}$~\cite{guo18}:
\begin{equation}\label{nacs}
d_{jk} = \frac{\left<\phi_j|\vec{\nabla}_{\!\!\scriptsize R} H|\phi_k\right>}{\epsilon_k-\epsilon_j}\frac{\partial\vec{R}}{\partial t},
\end{equation}
where $H$ is the electronic Hamiltonian. Evidently, NACs can enhance by (i) larger orbital overlaps, (ii) narrower energy separations, and (iii) faster nuclear velocities.

Fig.\,\ref{fig3} presents the trajectory-averaged magnitudes of NACs involving couplings among LUMO to LUMO+21 states in three xc frameworks employed. The diagonal trace shows zero signals due to the nonexistence of self-coupling. The super- and sub-diagonal traces represent predominant signals owing to the strongest couplings between nearest neighboring levels. Note that, universally in all three cases, the NAC signals between states LUMO+14 and LUMO+13, LUMO+6 and LUMO+5, LUMO+3 and LUMO+2 pairs are very weak, appearing practically dark in the color scale of Fig.\,\ref{fig3}. Obviously, this is due to the large energy difference, {\em via} the denominator of \eq{nacs}, from the gap between these states indicating their very weak mutual NAMD transition. Besides, as seen, there are differences in the finer details of NAC values among xc methods which have consequences in the dynamics they drive. We will address this in the next subsection.
%%%
\begin{figure}[h!]
%\centering
\includegraphics[width=9.0 cm]{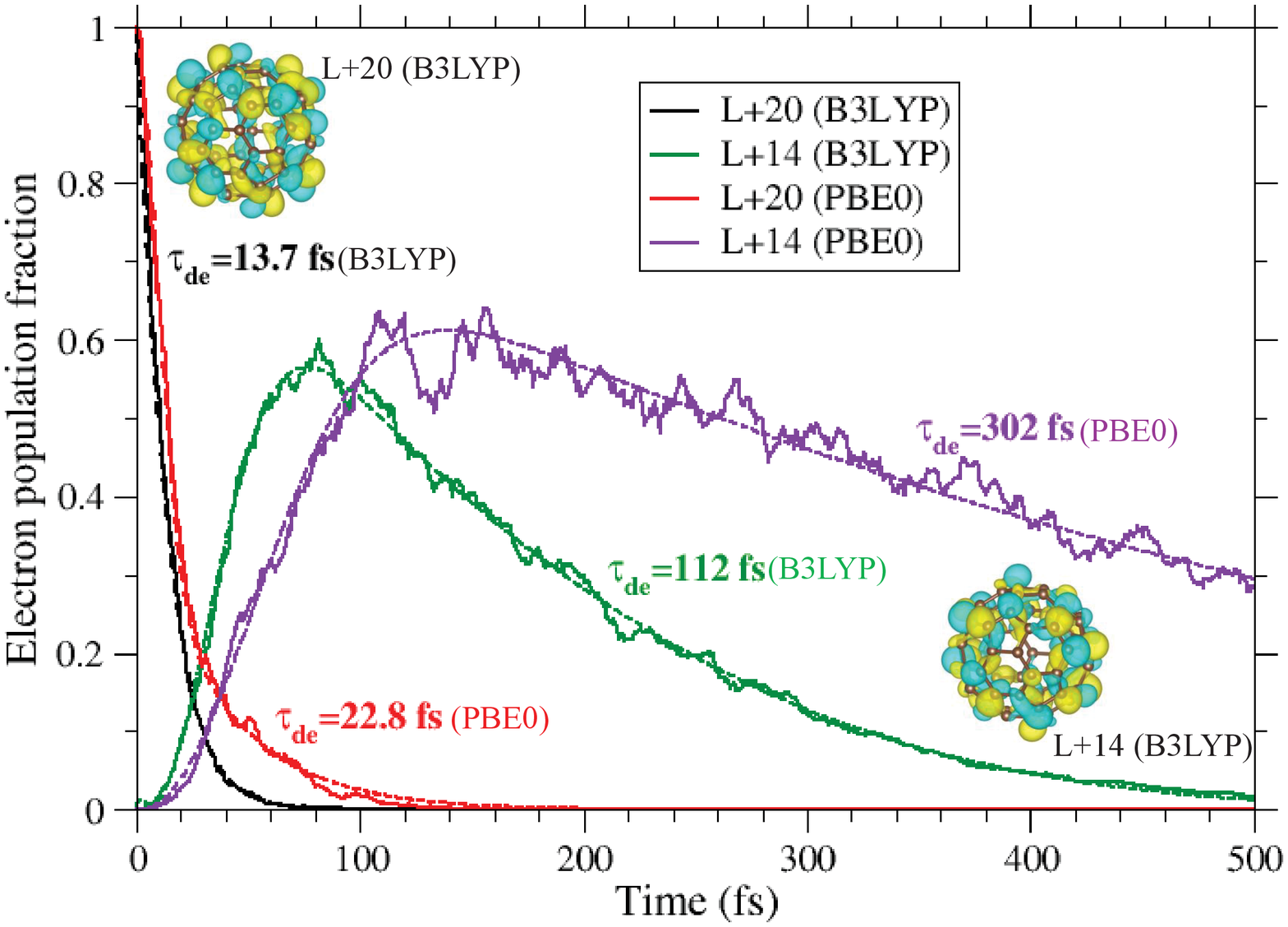}
%\vskip 1.5 cm
\includegraphics[width=9.0 cm]{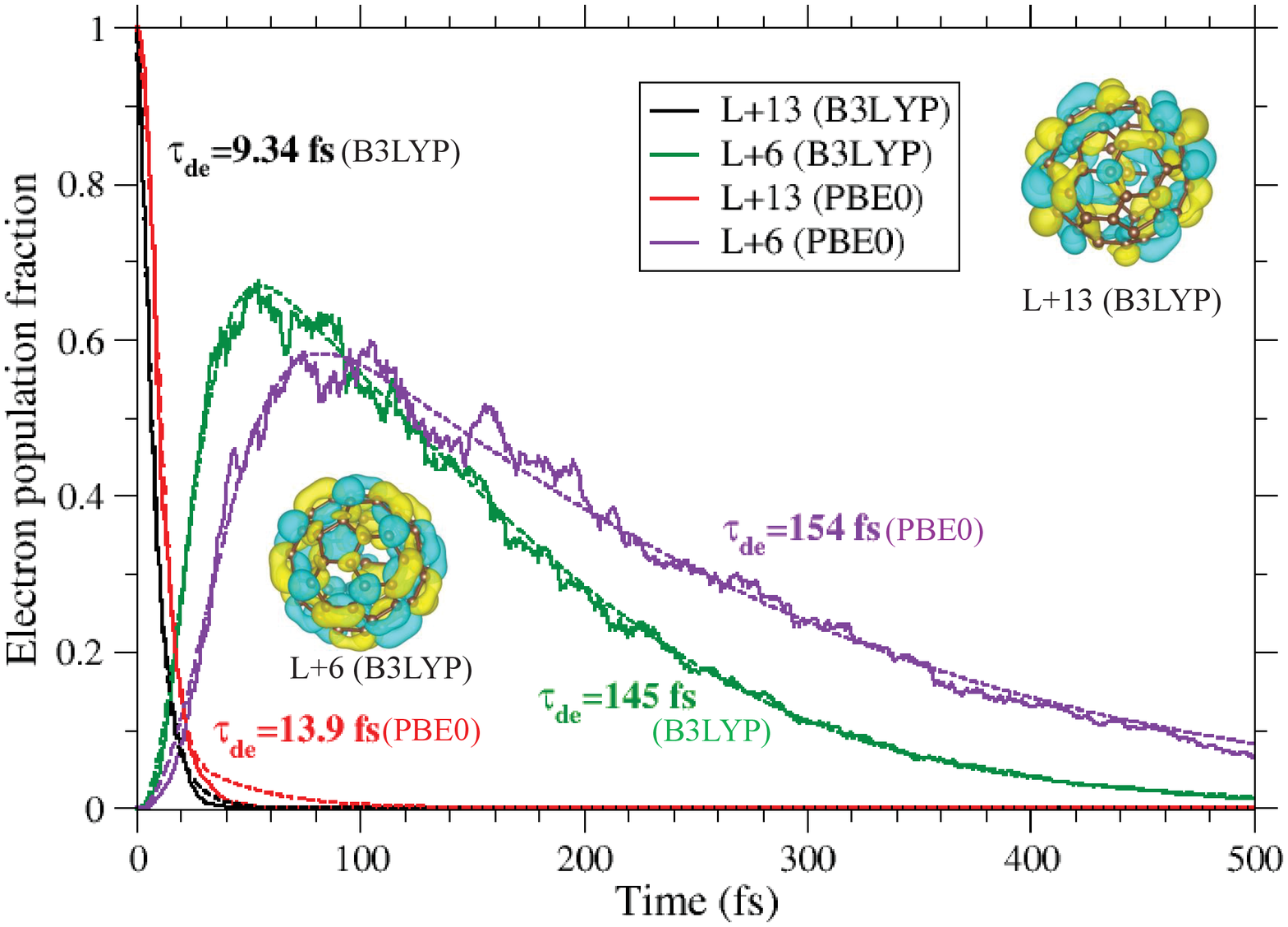}
\caption{Top: Time evolutions of the populations of initially excited LUMO+20 and the intermediate trapper state LUMO+14 calculated by B3LYP and PBE0.  The noted average decay times ($\tau_{de}$) are extracted by curve fittings (see text) and the fit curves for the decay are plotted by the dashed lines. The DFT/B3LYP generated isosurface orbital plots are illustrated. Bottom: Same as top, but for the initial LUMO+13 states and the trapper state LUMO+6.}
\label{fig4}
\end{figure}

\subsection{Decay and Transient-Capture Times}

We use the following fitting scheme to determine average evolution times~\cite{fisher14,madjet17}. The temporal decay of the population fraction of the initially excited state is fitted to the sum of an exponential and a Gaussian decay function appropriately weighted as, 
\begin{equation}\label{decay-fit}
F_{de} = a\exp(-t/a_1) + (1-a)\exp(-t^2/a_2^2),
\end{equation}
where $a$ is the weight parameter with a value between 0 and 1, and $a_1$ and $a_2$ are anti-steepness parameters such that the average decay time ($\tau_{de} $) is evaluated by 
\begin{equation}\label{time1}
\tau_{de} = aa_1+(1-a)a_2. 
\end{equation}
The evolution of the population fraction for an intermediate state, on the other hand, will involve a combination of both growth and decay processes. To stay consistent with \eq{decay-fit}, yet to introduce a different set of parameters, a fit formula for the growth component can be written as,
\begin{equation}\label{growth-fit}
F_{gr} = b[1-\exp(-t/b_1)] + (1-b)[1-\exp(-t^2/b_2^2)],
\end{equation}
with $0 < b < 1$. Thus, a general fitting formula for an intermediate state, affected by both decay and growth, can be considered as,
\begin{equation}\label{fit}
F = F_{gr} + F_{de} -1,
\end{equation}
with the average decay time 
\begin{equation}\label{time2}
\tau_{de} = bb_1+(1-b)b_2 + aa_1+(1-a)a_2.
\end{equation}

The NAMD simulation of the initially excited LUMO+20 state and the transient trapper state LUMO+14 is compared for B3LYP and PBE0 in Fig.\,4 (top panel). Fitting with \eq{decay-fit} produces 13.7 fs and 22.8 fs of average decay times [\eq{time1}] for LUMO+20 in, respectively, B3LYP and PBE0. An important mechanism that can influence a slower decay is the process of re-population of the state. This includes the return of electron population from the energetically lower states where the electron decayed into. The higher probability of repetition of this cycle will effectively sustain the net electron population longer. Obviously, the NACs govern these transition rates - the higher the value of NAC the stronger is the rate. As seen in Fig.\,\ref{fig3}, PBE0 NACs involving states from LUMO+20 and closely below are slightly stronger than those of B3LYP suggesting a more sustained back-and-forth of electrons producing longer decay time in PBE0. Furthermore, this difference is more pronounced for the decay of LUMO+14 that features, upon fitting with \eq{growth-fit}, an average time [\eq{time2}] of 112 fs in B3LYP and 302 fs in PBE0! Of course, a slightly larger gap below LUMO+14 in PBE0 (Fig.\,\ref{fig1}) favors this effect. But the more dominant reason of this difference is again the larger NAC values in PBE0 involving states LUMO+14 and above that ensures significant excitation to higher states in PBE0; this is qualitatively addressed earlier in the context of Fig.\,2 (top). This sputtered population to higher states keep feeding LUMO+14 back in order to facilitate its population to rise. This population maximizes until roughly 50 fs later in PBE0 than B3LYP, as seen. Therefore, stronger NACs in PBE0 result in LUMO+14's rather significantly longer decay time with this functional. 

Fig.\,4, bottom panel, explores the similar NAC-induced mechanism but for some lower states. This figure considers LUMO+13 as the initial excited state and also monitors population of the intermediate LUMO+6, which has another energy-gap below it (Fig.\,1). Maintaining the trend, LUMO+13 decays slower with an average of 13.9 fs in PBE0 than 9.34 fs in B3LYP. Evidently, this difference owes to the larger NAC values in PBE0, versus B3LYP, for states immediately below from LUMO+13 (see Fig.\,3). Remarkably, this trend reverses for NACs involving LUMO+6 and above, on the other hand, with PBE0 showing weaker values than B3LYP - an effect that reduces the degree of re-population of LUMO+6 in PBE0 than B3LYP. Indeed, the B3LYP peak in this case rises higher up to barely below 70\% even though still occurring 20 fs earlier. Consequently, the LUMO+6 average decay times produced by two functionals come very close being 145 fs (B3LYP) and 154 fs (PBE0). We note that the energy-gap below LUMO+6 is slightly larger in PBE0 that somewhat more hinders its decay across the gap.

A broad picture of the NAMD decay mechanism can now be drawn which is a cumulative account of three processes: (i) the direct population decay to energetically lower states (debit), (ii) the population back-transfer from those lower states (credit), and (iii) the return from the population promoted to energetically higher states (credit). For a state just above an energy-gap, like LUMO+14 and LUMO+6, (i) and (ii) above are very weak compared to (iii) leading to their significantly longer decay times. On the other hand, for a state just below a gap, say LUMO+13, the situation is quite the opposite with (iii) being too weak and (i) too strong resulting in their shortest decay times. Interestingly, for a state in the middle of an energy band, like LUMO+20, (iii) is not so weak with states above it within the band, while (i) is still strong. Consequently, the LUMO+20 decay-time will be slightly longer that LUMO+13, exactly as found and displayed in Fig.\,4. Hence, the state-selected ultrafast spectroscopy can access information that can map out some details of the LUMO structure.

We have noted, and as the trend of Fig.\,2 also suggests, the PBE derived dynamics (results not included) is generally fastest among three methods. This can be understood from the following observations in the PBE NACs in Fig.\,3. (i) The higher level PBE NACs involving LUMO+20 and the states below are (slightly) weaker than those of B3LYP. (ii) The NAC values for LUMO+13 and below in PBE exhibit the trend similar to B3LYP when compared to PBE0, although the high-end part in PBE is even weaker than B3LYP. (Incidentally, this general trend of NACs calculated in a non-hybrid functional like PBE being weaker than those obtained in hybrid functionals was suggested earlier~\cite{lin2016}). Moreover, as found, all the energy gaps in PBE are narrower than in other methods (Fig.\,1). All these have a net influence in reduced re-population resulting in a faster decay overall in PBE.  
%%%
\begin{figure}[h!]
%\centering
\includegraphics*[width=9.0 cm] {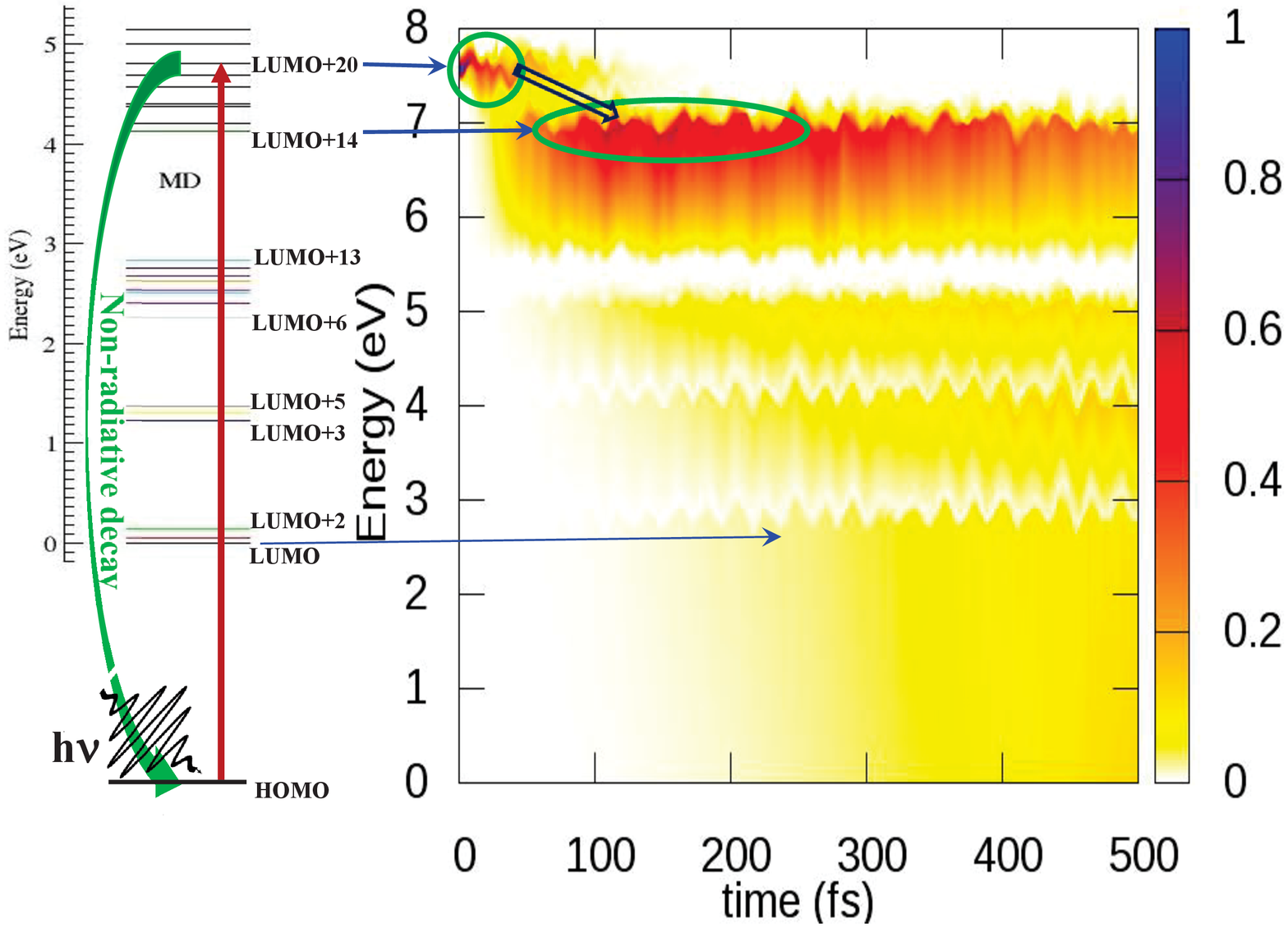}
%\vskip 1.5 cm
\includegraphics*[width=9.0 cm] {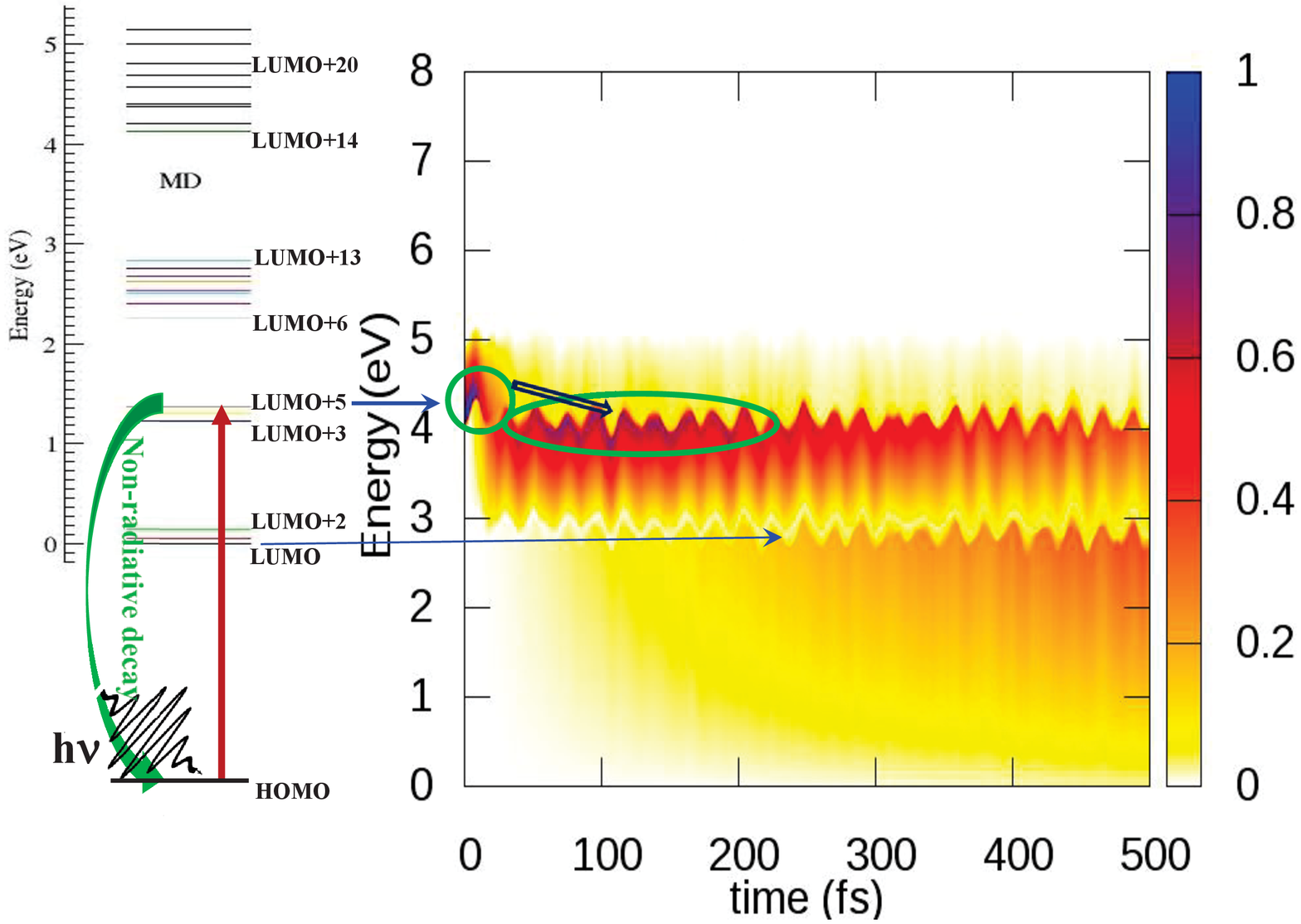}
\caption{(Color online) Top: The decay dynamics of initially excited LUMO+20 through all transient states producing a time-energy spectrogram contour map calculated in PBE0. The excited population decay of LUMO+20, transient capture in LUMO+14 and final recombination build-up in HOMO are indicated. Note that while in the LUMO energy band on the left the LUMO energy is set to zero, HOMO is considered zero on the spectrogram.  Bottom: Same as top, but for the initial excitation to LUMO+5 and transient capture in LUMO+3.} 
\label{fig5}
\end{figure}
%%%

\section{Spectrogram for Experiment}

As pointed out earlier, the ultrafast relaxation of pump laser-pulse excited $\ful$ molecules, in vapor or condensed matter phases, can be followed by a time-delayed probe pulse in both UTAS or TRPES schemes in experiment. The measurements can produce two-dimensional spectral information (spectrogram) as a function of the pump energy and pump-probe time delay~\cite{bhattacherjee2017,gesuele19,emmerich21,stadtmueller19}. Such spectrograms are directly comparable to the NAMD time-dependent population maps that our simulations can generate. Fig.\,\ref{fig5} includes such contour maps of the excited and transient electronic state population dynamics obtained from the DFT trajectories in our simulations for LUMO+20 and LUMO+5 initial pumped states on two panels. To illustrate, we only use results obtained by the PBE0 framework.

Fig.\,5, top panel, delineates transient electronic populations following the excitation from HOMO to LUMO+20 state by a far UV pump. From LUMO+20, the hot electron quickly decays through the states below to get transiently confined in LUMO+14 above the energy gap for a substantially extended time. The similar trend is qualitatively repeated for the states LUMO+6, LUMO+3, and LUMO, each top-edging a gap. However, note that a somewhat reduced  population is recorded going progressively further from LUMO+20 within the 500-fs time window shown. This slowdown toward the band edge is due to additional slowing effects induced during decay across multiple gaps. We remark that the oscillations in time noted in the contour plot are owing to lattice motions on the ground state potential energy surface coupled to the electron motion. Fig.\,5, bottom panel, exhibits a similar dynamics but for a mid-UV excitation to a lower LUMO+5 level. On both panels, the very light regions correspond extremely weak and fast decay through the compact energy band levels below the bottom-edge of the gaps. Also, some signal encroaching into the gaps is the numerical artifact due to the interpolation from strong population transiency at the gap-top. 

To summarize, within the reliability of a robust and ab\,initio methodology in DFT, the prediction of strong population traps atop the energy gaps on the decay path appears to be plausible and a fundamental effect. The fact that all these dynamics are reasonably captured as population growth and decay traces in Fig.\,5 bodes well for UTAS and/or TRPES measurements in accessing the structure of the fullerene excited states in general and probing dominant effects in particular. 

\section{Conclusion}

The ultrafast non-radiative relaxation process, driven by electron-phonon coupling (lattice thermalization), of a photoexcited electron in molecular $\ful$ is simulated in three exchange-correlation functionals, PBE, PBE0, B3LYP, within the DFT framework. The results are compared and analyzed which sheds light on the simultaneity of the decay, promotion and re-population processes from vibrionic motions in determining the dynamics. The study features a transient slowdown phenomenon, in the order of one hundred to a few hundreds of femtoseconds, of the relaxation process in real time due to the presence of large gaps in the spectrum of excited electronic states. The systematic trend of the decay at the gap bottom being fast, in orders of tens of femtoseconds, with progressive slowdown reaching up to the band top demonstrates a temporal route to tap in the excited state structural information by ultrafast spectroscopy. Ideally, the separation of quantum mechanical nuclear wave functions will reduce the coherence between electronic states~\cite{nelson2013,smith2019} -- a feature that is missing in our current semi-classical model of nuclear dynamics and is the topic of an upcoming study~\cite{wholey2023-damop}. It will further be interesting to know how the dynamics, in general, and the sizes of the LUMO gaps, in particular, quantitatively evolve in an electron-hole coupled-configurations framework. To this end, however, the effects found appear robust and fundamental, although the quantitative time information alters from one xc model to another, with PBE being the fastest and PBE0 the slowest but close to B3LYP.  

The study provides a motivation to conduct two-photon pump-probe UTAS or TRPES measurements on fullerene molecules, which are stable and symmetric and can be prepared in the vapor phase for experiments relatively easily. The measured spectrograms can be directly compared with the contour maps that can be simulated in our methods. The comparison will probe the current predictions. It will also help identify the best-performing xc scheme in order to optimize the computational methodology. Such optimization will be of great value to extend the study for the investigation of fullerene derivatives with increasing structural complications {\em via} endohedral / exohedral doping, functionalization and polymerization. We hope that the current results will pave experimental efforts in the domain of ultrafast science to complement our ongoing theoretical campaign.

\begin{acknowledgments} 
The research is supported by the National Science Foundation Grant Nos.\ PHY-1806206 and PHY-2110318, USA. M.E.M.\ acknowledges the German Research Foundation DFG (FR 2833/79-1) for financial support. Computing times at Bartik High-Performance Cluster (HPC) in Northwest Missouri State University (National Science Foundation Grant No.\ CNS-1624416, USA) and in Missouri University of Science and Technology, Rolla (National Science Foundational Grand No. OAC-1919789, USA) are acknowledged. 
\end{acknowledgments}

%\clearpage

\end{document}